\documentclass[aip,pof,amsmath,amssymb,preprint,linenumbers,
superscriptaddress,
]{revtex4-1}
\ifx\pdfoutput defined
\usepackage{graphicx}
\else
\usepackage{graphicx}
\usepackage{epstopdf}
\fi
\usepackage[pdftitle={POF},bookmarks=true,citecolor=red,colorlinks=false]{hyperref}
\usepackage{wasysym}
\usepackage{xcolor}

\usepackage{CJK}
\newcommand{\upd}{\mathrm{\,d}}

\begin{document}

  \linenumbers
%\setpagewiselinenumbers
\preprint{AIP/POF}
\begin{CJK*}{GB}{gbsn} % Use default fonts from CJK (see below)
\title[Scaling of pdf]{Scaling of maximum probability density functions of velocity and temperature increments in turbulent systems}
% Force line breaks with %\\

\author{Y.X. Huang (»ÆÓÀÏé)}
\email{yongxianghuang@gmail.com}
\affiliation{Shanghai Institute of Applied Mathematics and 
Mechanics,  Shanghai Key Laboratory of Mechanics in Energy and 
Environment Engineering, Shanghai University, Shanghai 200072, China
}
\affiliation{Modern Mechanics Division, E-Institutes of Shanghai Universities, 
Shanghai University, Shanghai 200072, China}
%\affiliation{Univ Lille Nord de France, F-59000 Lille, France}
%\affiliation{USTL, LOG, F-62930 Wimereux, France}
%\affiliation{CNRS, UMR 8187, F-62930 Wimereux, France}
%Lines break automatically
%or can be forced with \\

\author{F. G. Schmitt}%
\email{francois.schmitt@univ-lille1.fr}
\affiliation{Universit\'e Lille Nord de France, F-59000 Lille, France}
\affiliation{USTL, LOG, F-62930 Wimereux, France}
\affiliation{CNRS, UMR 8187, F-62930 Wimereux, France}

\author{Q. Zhou (ÖÜÈ«)}
\affiliation{Shanghai Institute of Applied Mathematics and 
Mechanics,  Shanghai Key Laboratory of Mechanics in Energy and 
Environment Engineering, Shanghai University, Shanghai 200072, China
}
\affiliation{Modern Mechanics Division, E-Institutes of Shanghai Universities, 
Shanghai University, Shanghai 200072, China}

\author{X. Qiu (ÇñÏè)}
\affiliation{School of Science, Shanghai Institute of Technology, 200235, Shanghai, China}%
% \affiliation{Univ Lille Nord de France, F-59000 Lille, France}
% \affiliation{USTL, LML, F-59655 Villeneuve d'Ascq, France}
%\affiliation{CNRS, UMR 8107, F-59655 Villeneuve d'Ascq, France}
% \affiliation{USTL, LOG, F-62930 Wimereux, France}

\author{X.D. Shang (ÉÐÏþ¶«)}
\affiliation{State Key Laboratory of Tropical Oceanography£¬South China Sea Institute of Oceanology,Chinese Academy of Sciences, Guangzhou 510301, China}%

 \author{Z.M. Lu (¬־Ã÷) }  
\affiliation{Shanghai Institute of Applied Mathematics and 
Mechanics,  Shanghai Key Laboratory of Mechanics in Energy and 
Environment Engineering, Shanghai University, Shanghai 200072, China
}
\affiliation{Modern Mechanics Division, E-Institutes of Shanghai Universities, 
Shanghai University, Shanghai 200072, China}

\author{Y.L. Liu (ÁõÓî½)}
\affiliation{Shanghai Institute of Applied Mathematics and 
Mechanics,  Shanghai Key Laboratory of Mechanics in Energy and 
Environment Engineering, Shanghai University, Shanghai 200072, China
}
%\affiliation{Modern Mechanics Division, E-Institutes of Shanghai Universities, 
%Shanghai University, Shanghai 200072, China}

\date{\today}% It is always \today, today,

\begin{abstract}
In this paper,  we introduce a new way to estimate the scaling parameter of a self-similar process by considering the maximum probability density function (pdf) of tis increments. We prove this for $H$-self-similar processes in general and experimentally investigate it for turbulent
 velocity and temperature increments.   We consider  turbulent velocity database from an experimental
homogeneous and nearly isotropic turbulent channel flow, and  temperature data
set  obtained near the sidewall of a Rayleigh-B\'{e}nard convection cell, where
the turbulent flow is driven by buoyancy. For the former database, it is found  that  the
maximum value of increment pdf $p_{\max}(\tau)$ is in a good agreement with
 lognormal distribution.  We also obtain a scaling exponent
$\alpha\simeq 0.37$, which is consistent with the scaling exponent for the first-order structure function reported in other studies. For the latter one, we
obtain a scaling exponent $\alpha_{\theta}\simeq0.33$. This index value is  consistent
with the Kolmogorov-Obukhov-Corrsin scaling for passive scalar turbulence, but different from the scaling exponent of the
first-order structure function that is found to be $\zeta_{\theta}(1)\simeq
0.19$, which is in favor of Bolgiano-Obukhov scaling. A possible explanation for these results is also given.
\end{abstract}

%{Time series analysis}
%\pacs{02.50.Fz}{Stochastic analysis}
%\pacs{47.27.Gs}{Isotropic turbulence; homogeneous
%turbulence}
%\pacs{47.53.+n}{Fractals in fluid dynamics}
%\keywords{autocorrelation function, power law}%Use showkeys class option if keyword
\maketitle
\end{CJK*}

\section{Introduction}

%\mysout{delete something} \red{add something}

Since Kolmogorov's 1941 (K41) milestone work, the invariant properties of small-scale
structures have been widely investigated during the last four decades.\cite{Kolmogorov1941,Anselmet1984,Frisch1995,Sreenivasan1997,Lohse2010}
The invariant properties are characterized by a series of scaling exponents $\zeta(q)$, which is
traditionally extracted by the classical structure function  (SF) analysis $S_q(\ell)=\langle
\Delta u_{\ell}(r)^q\rangle\sim \ell^{\zeta(q)}$ that has been documented very well for turbulent
velocity fields.\cite{Anselmet1984,Frisch1995,Arneodo1996,Sreenivasan1997} Here, $\Delta u_{\ell}(r)=u(\ell+r)-u(r)$ is the velocity increment.

A key problem of turbulence is the search of universal probability density function (pdf) of turbulent velocity.\cite{Monin1971,Frisch1995} The pdf of turbulent velocity or velocity increments has been studied by several authors.\cite{Anselmet1984,Castaing1990,Ching1991PRA,Kailasnath1992PRL,Ching1993PRL,Tabeling1996PRE,Noullez1997JFM}
Several models of velocity increment have been proposed to characterize its pdf tail. For example, \citet{Anselmet1984} proposed an exponential fitting to extrapolate the pdf tail of velocity increments with separation scales in inertial range, which is also advocated in Ref. \onlinecite{Castaing1990}.  \citet{Ching1991PRA} proposed a stretched exponential pdf of temperature increments for Rayleigh-B\'enard convection  (RBC) system. Later, it has been applied in turbulent velocity by \citet{Kailasnath1992PRL}.

In this paper, we investigate another aspect of the pdf scaling of increments of  scaling
time series, e.g. fractional Brownian motion (fBm), turbulent velocity, and temperature. 
 We find a pdf scaling 
\begin{equation}
p_{\max}(\tau)\sim \tau^{-\alpha}\label{eq:PDFS}
\end{equation} 
for the maximum pdf  of  the
increment $\Delta u_{\tau}$. This pdf scaling can be obtained analytically for the fBm processes and more 
generally for $H$-self-similar processes, in which only one parameter Hurst number $H$ 
is required to describe the processes, and we find in this case $\alpha=H$. For the fBm case, the pdf scaling is validated by 
numerical simulations. We hence postulate that the pdf scaling also holds for multifractal 
processes, such as turbulent velocity, temperature fluctuations in RBC system, etc. 
To our knowledge, the method we proposed here is the first method to extract scaling exponents on the probability space rather than on 
the statistical moments space, as usually done. 

This paper is organized as following. In section \ref{sec:fBm},  we derive analytically a pdf scaling of 
increments for  fractional Brownian motion processes  and more generally for $H$-self-similar processes. In section \ref{sec:ER},  we  investigate the pdf scaling
of velocity from turbulent channel flow and temperature from turbulent Rayleigh-B\'ernard convection system, respectively. We finally present
our discussions and draw the main conclusion in
section \ref{sec:DCs}.

\section{fractional Brownian motion and $H$-self-similar processes }\label{sec:fBm}

\begin{figure}[!htb]
\centering
\includegraphics[width=0.95\linewidth]{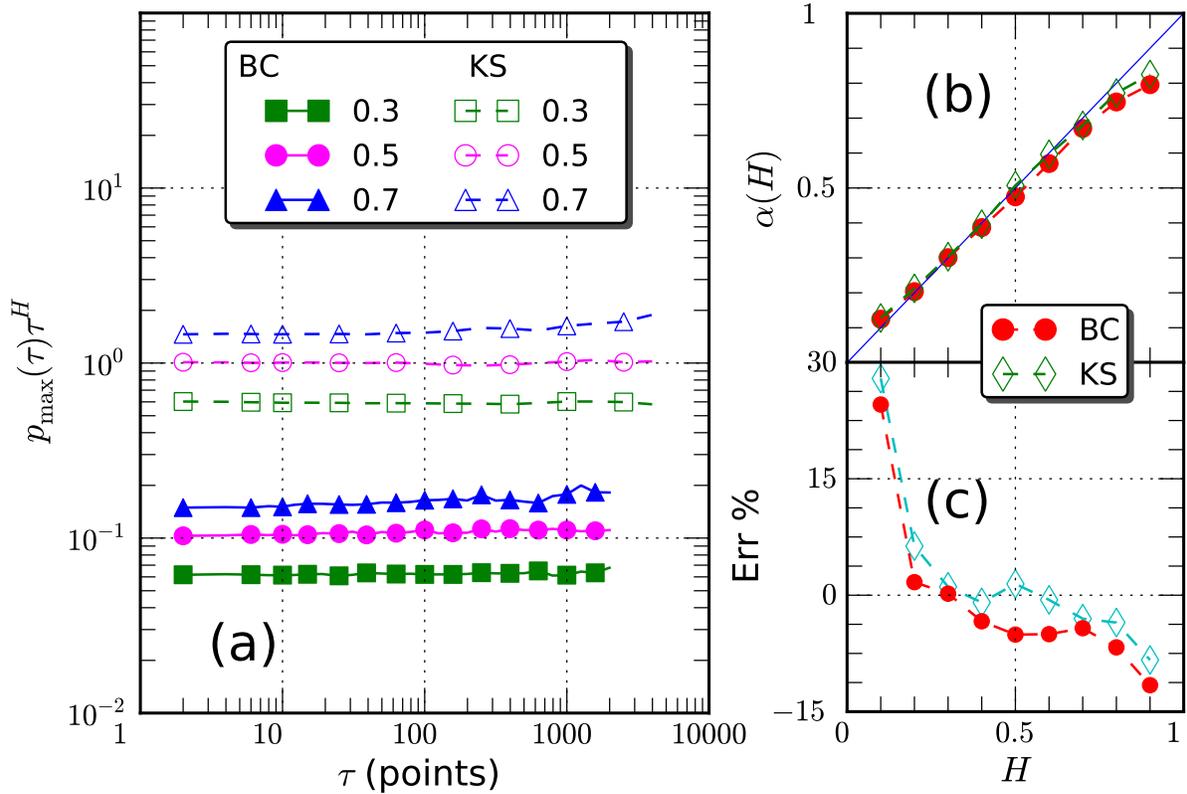}
\caption{ (a)  Compensated $p_{\max}(\tau)\tau^{H}$ estimated from fBm simulation with various
Hurst numbers $H$ by using box-counting method (denoted as BC) and a kernel smoothing method  with Gaussian kernel (denoted as KS), (b) the corresponding scaling exponents
$\alpha(H)$, where the theoretical value is illustrated by a solid line, and (c) the relative error  $(\alpha-H)/H$  between given and estimated Hurst number. The
scaling exponent is estimated on the range $10<\tau<1000$ data points by using a
least square fitting algorithm.}\label{fig:fBm}
\end{figure}

\begin{figure}[!htb]
\centering
\includegraphics[width=0.95\linewidth]{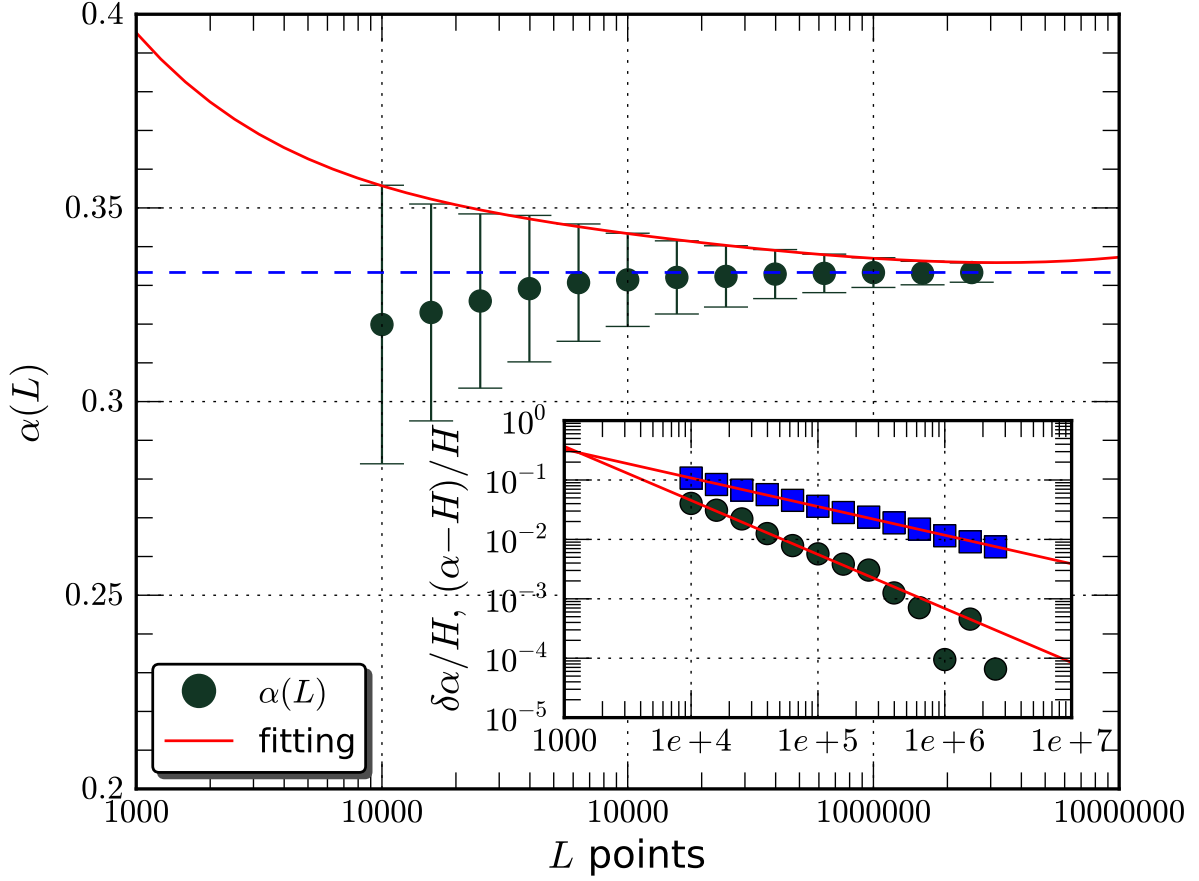}
\caption{ The length dependent $\alpha(L)$ ($\ocircle$) for the Hurst number $H=1/3$ with 1000 realizations (using box-counting method). The horizontal 
dashed line illustrates the given Hurst number $H=1/3$. The solid line is the fitting of the errorbar (standard deviation of $\alpha(L)$). The inset shows the corresponding relative error $Er(L)=(H-\alpha(L))/H$ ($\ocircle$) and errorbar $\delta \alpha(L)/H$ ($\square$), in which the solid line is the power law fitting $Er(L)\sim L^{-\xi}$ and $\delta \alpha(L)\sim L^{-\gamma}$ with scaling exponents  $\xi\simeq 0.91$ and $\gamma\simeq 0.49$ respectively.  }\label{fig:length}
\end{figure}

\subsection{Fractional Brownian motion}

FBm  is a continuous-time random process proposed
by \citet{Kolmogorov1940} in the 1940s and \citet{Yaglom1957} and later named `fractional
Brownian motion'
by Mandelbrot \cite{Mandelbrot1968}. It consists in a fractional integration of a white
Gaussian process and is therefore a generalization of Brownian motion, which
consists simply in a standard integration of a white Gaussian
process. Because it presents
deep connections with the concepts of self-similarity, fractal, long-range
dependence or $1/f$-process, it quickly became a major tool for  various
fields where such concepts are relevant, such as in geophysics, hydrology,
turbulence, economics, communications, etc. \cite{Mandelbrot1968,Flandrin1992,Samorodnitsky1994,Beran1994,
Rogers1997,Doukhan2003,Gardiner2004,Biagini2008,Huang2009EPL}
Below we consider  it as an analytical model for monofractal processes to obtain the pdf scaling analytically.

An autocorrelation  function of fBm's increments $Y_{\tau}(t)=x(t+\tau)-x(t)$ is known
to
be the following  
\begin{equation}
R_{\tau}(\ell)=\frac{1}{2}\{ (\tau+\ell)^{2H}+\vert \tau-\ell
 \vert^{2H}-\ell^{2H} \}
\end{equation}
where $\ell\ge0$ is the time delay, $\tau$ is the separation scale, and $H$ is Hurst
number. \cite{Biagini2008}
Thus the standard deviation $\sigma(Y_\tau)$ of the increment $Y_{\tau}$  
scales as
\begin{equation}
\sigma(Y_\tau)=R_{\tau}(0)^{1/2}=\tau^{H}
\end{equation}
  $Y_{\tau}$  is also known to have a Gaussian distribution,
\cite{Mandelbrot1968,Biagini2008}  which reads as
\begin{equation}
p(Y_{\tau} )=\frac{1}{\sigma(Y_\tau)\sqrt{2\pi}}\exp\left(
-\frac{Y_{\tau}^2}{2\sigma(Y_\tau)^2} \right)
\end{equation}
We thus have a power law relation when $Y_{\tau}=0$
\begin{equation}
p_{\max}(\tau)=p(Y_\tau)\vert_{Y_{\tau}=0} =\frac{1}{\sigma(Y_\tau)\sqrt{2\pi}}%=\frac{1}{\sigma(\tau)\sqrt{2\pi}}
=\frac{1}{\sqrt{2\pi}} \tau^{-\alpha( H)}\label{eq:pl2}
\end{equation}
where $\alpha( H)=H$.

In order to numerically check this, we  perform  a wavelet based algorithm to simulate the fBm process.
\cite{Abry1996}
We synthesize a segment of length $10^6$ data points  for each value of Hurst number $H$ from 0.1 to 0.9 by using
db2 wavelet. The pdfs are estimated as  follows.  We first normalize $x$ by its  own standard deviation $\sigma$.
The empirical pdf  is then estimated by using box-counting method on several discrete bins with width $\upd h$
\begin{equation}
p(Y)=\frac{N_i}{N\upd h}\label{eq:pdf}
\end{equation}
in which $N_i$ is the number of events in the $i$th bin, $N$ is the total length of the data.
    We find that the  empirical pdf $p(Y_{\tau})$, the maxima pdf $p_{\max}(\tau)=\max_{Y_{\tau}}\left\{ p(Y_{\tau}\vert \tau) \right\}$, and the corresponding scaling
exponents $\alpha(H)$  are almost independent of the  range of bin width $\upd h$.  Another way to estimate  the pdf is a kernel smoothing   method. \cite{Wand1995}  In this study, a Gaussian kernel is chosen.
 Figure \ref{fig:fBm} (a) shows the estimated  $p_{\max}(\tau)\tau^{H}$ 
for various Hurst numbers $H$. For both methods,  a clear plateau is observed, indicating power law behavior  as expected for
all Hurst numbers. We  estimate the scaling exponents on the range
$10<\tau<1000$ data points by a least square fitting
algorithm. The corresponding scaling exponents  $\alpha(H)$ are shown in Fig.\,\ref{fig:fBm} (b).
One can see that, except for the larger   values of $H$, the scaling exponents $\alpha(H)$
 are in good agreement with the given Hurst numbers.
 We note that our new method overestimated $H$ for small values of $H$, and then underestimated it  for high values.  
    We then check the relative error $(\alpha(H)-H)/H$ between given and estimated Hurst number.  The corresponding result 
is shown in Fig.\,\ref{fig:fBm} (c).  Generally speaking, both the kernel smoothing method and the box-counting method provide a comparable estimation of $H$, especially for a Hurst number around $H=0.3$. Thus in the following content, we will only apply the 
box-counting method to real data sets since the scaling exponent is expected around $H=1/3$.
     
 To   test the  finite length effect, we perform a calculation with various data length 
$L$ and 1000 realizations each for the Hurst number $H=1/3$, this corresponds to the value for  fully developed turbulence.
\cite{Frisch1995} The range of $L$ is $10^4<L<3\times 10^6$. The corresponding scaling 
exponents $\alpha(L)$ are estimated on the range $10<\tau<1000$ data points. Figure 
\ref{fig:length} shows    $\alpha(L)$ with errorbar $\delta \alpha(L)$, which is the 
standard deviation of the estimated $\alpha(L)$.  The inset shows the 
relative error $Er(L)=(H-\alpha(L))/H$ ($\ocircle$) and the  errorbar $\delta \alpha (L)/H$ ($\square$). Power law behaviors
\begin{equation}
Er(L)\sim (L)^{-\xi},\quad\delta \alpha (L)\sim 
L^{-\gamma}
\end{equation} are observed with scaling exponents  $\xi\simeq 0.91$ and $\gamma 
\simeq 0.49$. One can find that the estimated $\alpha$ is quickly close to the 
given Hurst number $H=1/3$. The relative error $Er(L)$ is less than $10\%$ for 
all $L$ we considered here. Specifically, when $L>10^5$, we obtain $Er(L)\le 1\%$ and $\delta \alpha (L)/H\le 5\%$. This is already a quite good estimation of $H$. Thus in the following, we choose $L\ge 10^5$ data points.

\subsection{$H$-Self-similar processes}

We can also derive the pdf scaling more generally for $H$-self-similar processes as following.
We define a $H$-self-similar process as 
\begin{equation}
\left\{     x(at) \right\}  \stackrel{d}{=} \left\{  a^H x(t)  \right\} \label{eq:PR0}
\end{equation}
in which $ \stackrel{d}{=}$ means equality in distribution and  $H$ is the Hurst number.\cite{Embrechts2002selfsimilar} $x(t)$ is a $H$-self-similar process, in which only one parameter $H$, namely Hurst number, is required for the above scaling transform.
Let us note $Y_{\tau}=\Delta x_\tau =x(t+\tau)-x(t)$, the increment with separation scale $\tau$. We assume that $x$ is $H$-self-similar with stationary increment, hence $Y_{\tau}$ is also $H$-self-similar.
Thus one has
\begin{equation}
\left\{ \frac{Y_{\tau}}{\tau^H} \right\}  \stackrel{d}{=} \left\{ \frac{Y_{T}}{T^H} \right\}  \stackrel{d}{=} \left\{  Y_{1}\right\} \label{eq:PR}
\end{equation}
In fact equality in distribution means equality for distribution function. Let us write distribution function 
\begin{equation}
F(x)=P_r(X\le x)=\int_{-\infty}^{x}p(X)\upd X
\end{equation}
in which $p(x)$ is the pdf of $x$. We note the pdf
\begin{equation}
p(x)=F'(x)
\end{equation} 
We thus take here 
\begin{equation}
F_{\tau}(x)=P_r\left( Y_{\tau}\le x \right)
\end{equation}
We have 
\begin{equation}
P_r\left(  \frac{Y_{\tau}}{\tau^H}  \le x\right)=P_r\left( Y_{\tau}\le x \tau^H \right)
\end{equation}
Hence Eq.\,\eqref{eq:PR} writes for distribution functions
\begin{equation}
F_{\tau}\left( x\tau^H \right)=F_T\left( x T^H \right)\label{eq:PR2}
\end{equation}
Taking the derivative of Eq.\,\eqref{eq:PR2}, we have for the pdfs
\begin{equation}
\tau^Hp_{\tau}\left( x\tau^H \right)=T^Hp_{T}\left( x T^H \right)\label{eq:PR3}
\end{equation}
Then writing 
\begin{equation}
p_{\max}(\tau)=\max_{x}\left\{ p_{\tau}(x) \right\}  
\end{equation}
and taking the maximum of Eq.\,\eqref{eq:PR3}, we have
\begin{equation}
\tau^Hp_{\max}(\tau)=T^Hp_{\max}(T)\label{eq:PR4}
\end{equation}
Finally, this leads to
\begin{equation}
p_{\max}(\tau)=p_{\max}(T)\left(
\tau/T%\frac{\tau}{T}
\right)^{-H}\label{eq:PR5}
\end{equation}
This is the pdf scaling for the $H$-self-similar process.
Since Eq.\,\eqref{eq:PR0} is not true for multi-scaling processes, Eq.\,\eqref{eq:PR5} may be only an approximation for multifractal processes.

  We have shown above analytically  the pdf scaling relation  
for fBm processes and more generally for $H$-self-similar processes. For the former one, the pdf scaling Eq.\,\eqref{eq:PDFS} is validated by numerical simulations. We postulate here that it is 
 also valid for other types scaling time series, e.g. turbulent velocity and temperature from other turbulent systems, etc.,  and we will check this experimentally in the next section.

 The pdf scaling we proposed above is related to the first-order structure function $\alpha=H$ for $H$-self-similar
  processes, see Eqs.\,\eqref{eq:pl2} and \eqref{eq:PR5}. 
 Hence for the multifractal case, we may postulate that $\alpha=\zeta(1)$, the-first order structure function with a slight intermittent correction, see next section for turbulent velocity as an example. 

\section{Experimental results}\label{sec:ER}
In this section, we will  apply the above pdf scaling analysis to turbulent velocity obtained from homogeneous and nearly isotropic channel flow, and temperature time series obtained near sidewall area of a Rayleigh-B\'enard convection system. An interpretation under the K41 theory is also discussed.

\subsection{Turbulent velocity}

 \begin{figure}[!htb]
\centering
\includegraphics[width=0.95\linewidth]{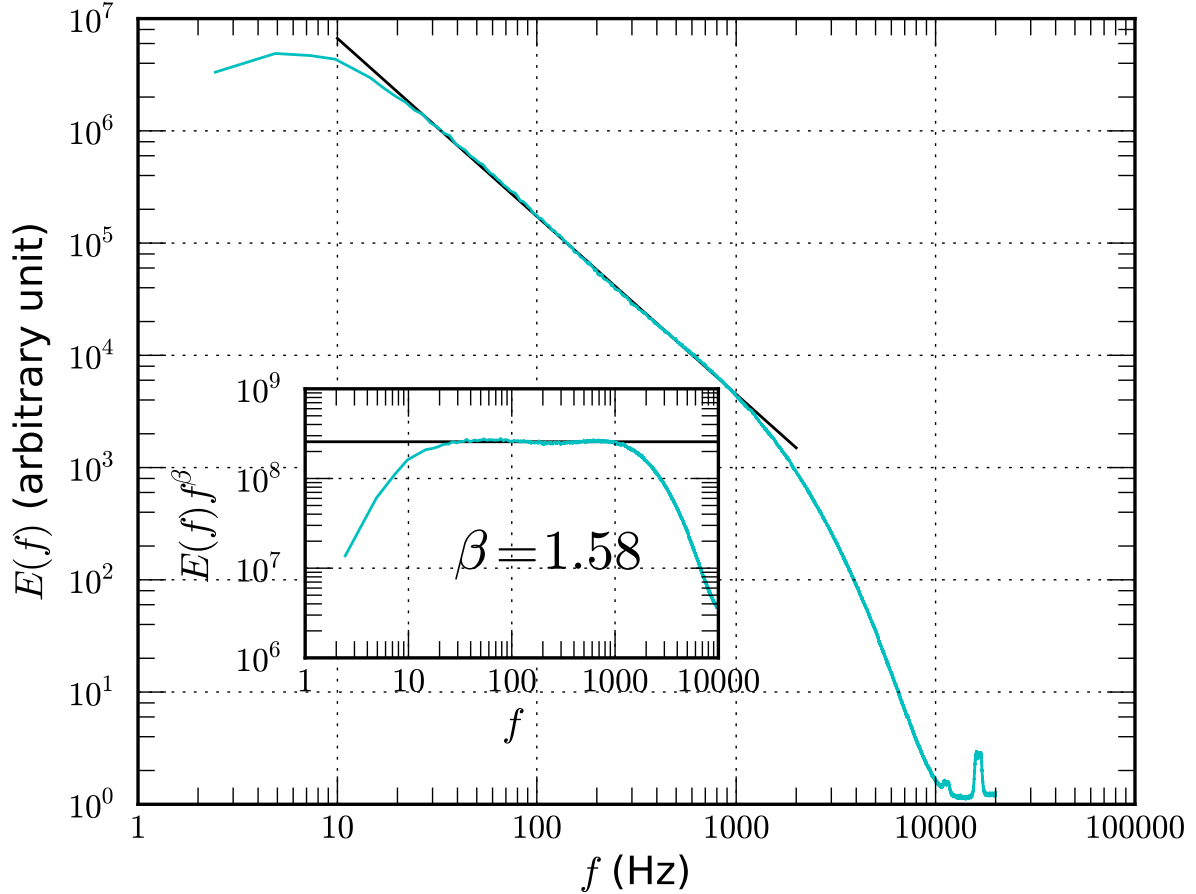}
\caption{ Fourier power spectrum of
transverse velocity component at downstream $x/M=48$, where $M$ is the mesh
size. The inset shows the corresponding compensated
spectra $E(f)f^{\beta}$, where $\beta\simeq1.58$ is estimated on the range
$20<f<1000\,$Hz, corresponding to the time separation $\tau$ on the range
$0.001<\tau<0.05\,$s. }\label{fig:psdv}
\end{figure}

 \begin{figure}[!htb]
\centering
\includegraphics[width=0.95\linewidth]{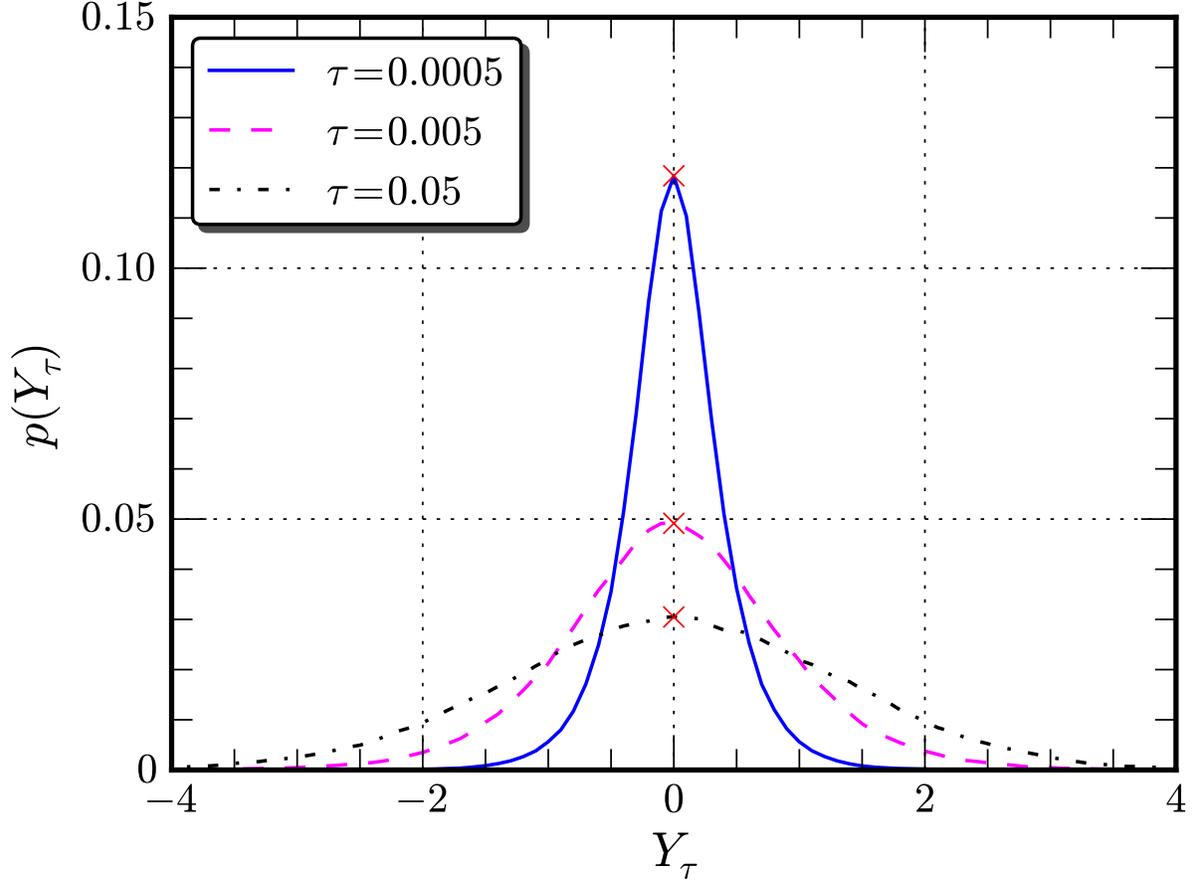}
\caption{Empirical pdf estimated for transverse velocity with time separation $\tau=0.0005,\,0.005$, and $0.05\,$s, corresponding to $f=2000,\,200$ and $20\,$Hz. The location for $p_{\max}(\tau)$ is marked by $\times$.}\label{fig:pdfv}
\end{figure}

 \begin{figure*}
\begin{center}
\includegraphics[width=.7\linewidth]{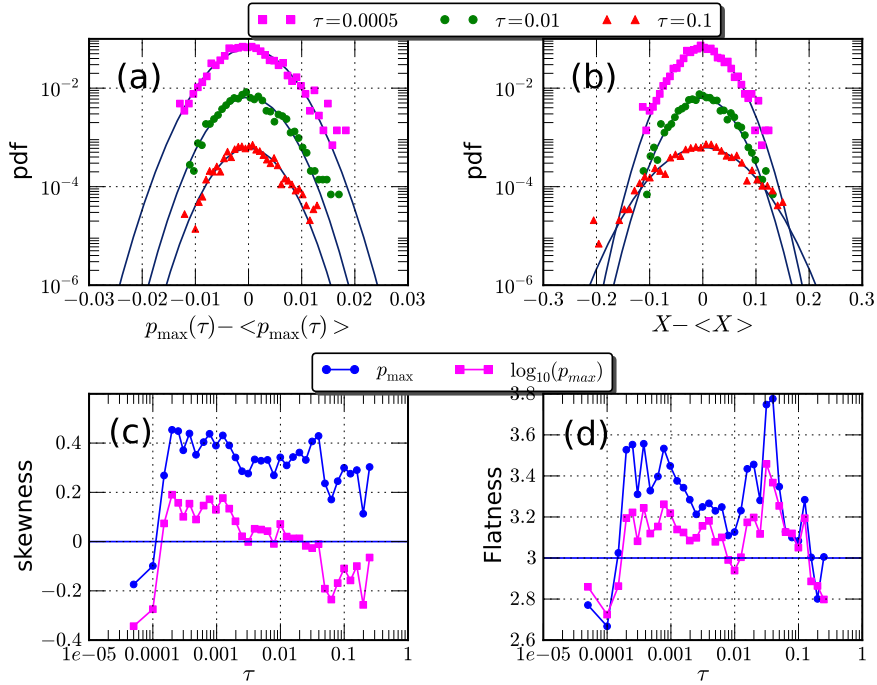}
\caption{(a) pdf of $p_{\max}(\tau)$,  (b) pdf of
$X=\log_{10}(p_{\max}(\tau))$, and (c) the
skewness factor of $p_{\max}$ (\newmoon) and $\log_{10}(p_{\max})$, and (d) the flatness factor.  Separation scales  in (a) and (b)
are $\tau=0.0005$ s ($\blacksquare$), $0.01$ s ($\newmoon$) and $\tau=0.1\,$s
($\blacktriangle$),
corresponding to $f=2000$ Hz in dissipation range, $100$ Hz in inertial range, and
$10$ Hz in large scale forcing range, respectively. For
display convenience in (a) and (b), the mean value of each scale is removed and
the curves have
been vertically shifted. The normal distribution is illustrated by  a solid line.
}\label{fig:lognormalpdfv}
\end{center}
\end{figure*}

 \begin{figure}[!htb]
\begin{center}
\includegraphics[width=.95\linewidth]{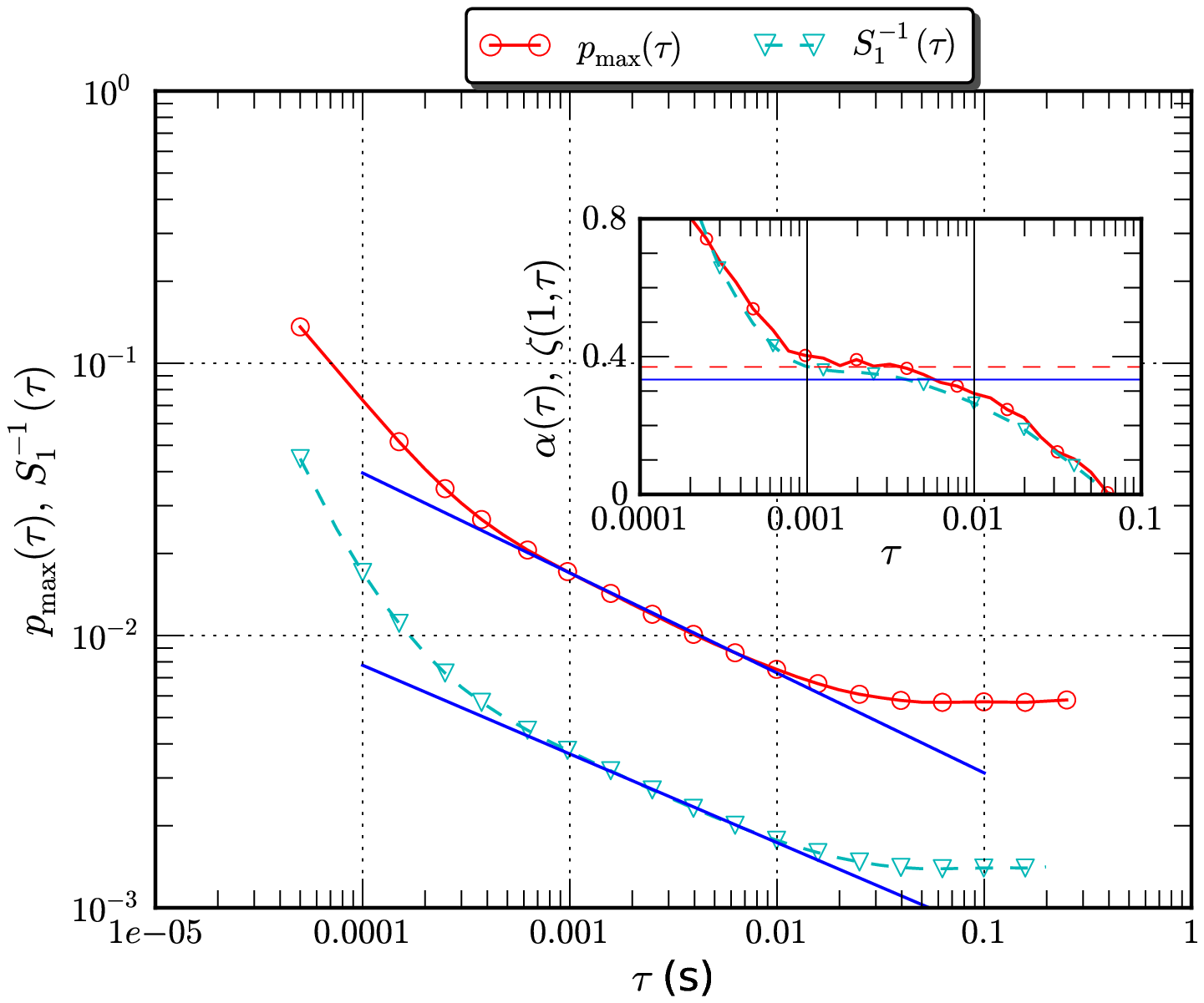}
\caption{    $p_{\max}(\tau)$  and the first-order structure function $S_1(\tau)$ of transverse   velocity
component.
 The scaling exponents are $\alpha\simeq 0.37$  and
$\zeta(1)\simeq 0.34$   estimated on the power law range
$0.001<\tau<0.01\,$s, corresponding
to $100<f<1000\,$Hz.  The inset shows the local slope $\alpha(\tau)$ and $\zeta(1,\tau)$, in which the horizontal solid line indicates the Kolmogorov value $1/3$, and the dashed line 
indicates the value $0.37$, and the vertical solid line illustrates the inertial range.
For display convenience, we have shown the inverse value $1/S_1(\tau)$ of the first-order structure function and the curves have been vertically shifted.  }\label{fig:maxpdfv}
\end{center}
\end{figure}

We  consider here a turbulent
velocity database obtained from an experimental homogenous and nearly isotropic
turbulent channel flow  by using an active-grid technique to achieve a high Reynolds number.\cite{Kang2003}   We use the data obtained at
downstream $x/M=48$, where $M$ is the mesh size. At this measurement location,
the mean velocity is $\langle u
 \rangle=10.8\,$m/s, the turbulence intensity is 10\%, and the Taylor
microscale based Reynolds number is $Re_{\lambda}\simeq 630$. The sampling
frequency is $40,000\,$Hz.    To avoid the
measurement noise, we only consider here the transverse velocity.
 Figure \ref{fig:psdv} shows the Fourier power spectrum for  the transverse velocity
component.  The inset shows the compensated spectrum $E(f)f^{\beta}$, in which
$\beta\simeq 1.58$ is the scaling exponent estimated on the range $20<f<1000\,$Hz,
corresponding to the time separation $0.001<\tau<0.05\,$s. The value of $\beta$ for transverse velocity component at all measurement locations ($x/M=20,\,30,\,40,\,48$) is around $1.58\sim 1.60$, and is slightly smaller than the Kolmogorov value $5/3$, which could be an effect of the active-grid technique.
\footnote{However, for the second-order structure functions, the corresponding scaling exponent is $\zeta(2)\simeq 0.64$ (the figure not shown here). It indicates that the relation $\beta=1+\zeta(2)$ does not hold, which has been understood as an effect of large-scale structure and finite scaling range. Another example for passive scalar has been shown in Ref. \onlinecite{Huang2010PRE}.}  It
demonstrates a nearly two decades inertial range.  Thus this database has
a long enough
 inertial
range to validate Eq.\,\eqref{eq:PDFS}.  More details about this database can
be
found in Ref. \onlinecite{Kang2003}.

$p_{\max}(\tau)$ is calculated as  explained below. We first divided the time series
into
several segments with $10^5$ data points each. Then empirical pdf is estimated for various separation scales by using Eq.\,\eqref{eq:pdf}. $p_{\max}(\tau)$ is then
estimated for each segment. We  have $120\times 12=1440$
 (number of measurements $\times$ segments of each measurement) realizations.
Figure \ref{fig:pdfv} shows the estimated pdf for several separation scales $\tau$ for one realization.  
The location of  maxima pdf $p_{\max}(\tau)$ is marked by $\times$. Graphically, $p_{\max}(\tau)$ decreases with $\tau$ and the corresponding location is around, not exactly, $Y_{\tau}=0$.
Figure \ref{fig:lognormalpdfv} shows (a) the pdf of  $p_{\max}$ for several separation scales $\tau$, (b)
$X=\log_{10}(p_{\max})$,  (c) the skewness factor of $p_{\max}$ and $\log_{10}(p_{\max})$, and (d) the the flatness factor. The separation scales 
in Fig.\,\ref{fig:lognormalpdfv} (a) and (b)  are $\tau=0.0005$
($\blacksquare$), $0.01$
($\newmoon$) and $\tau=0.1\,$s ($\blacktriangle$),
corresponding to $f=2000\,$Hz in dissipation range, $100\,$Hz in inertial range
and
$10\,$Hz in large scale forcing range, respectively.
 Both normal and lognormal fits seem to capture the fluctuations of $p_{\max}(\tau)$. It seems that the lognormal fit is better than the normal one, but more data are certainly needed to remove measurement uncertainty and to determine  without ambiguity which pdf fit is closest to the data.

Figure \ref{fig:maxpdfv} shows the ensemble averaged $p_{\max}(\tau)$ for the
transverse
velocity ($\ocircle$).  The inset shows the local slope, in which the horizontal solid line illustrates the Kolmogorov value $1/3$ and the dashed line illustrates the value $0.37$, and the vertical solid line demonstrates the plateau range, e.g. the inertial range $0.001<\tau<0.01\,$s.
Here 
the local slope is defined as
\begin{equation}
\alpha(\tau)=-\frac{\upd \log_{10}(p_{\max}(\tau))}{\upd \log_{10}(\tau)}
\end{equation}
 A power law  behavior is observed over the range $0.001<\tau<0.01\,$s,
corresponding
to the frequency range $100<f<1000\,$Hz.   
 This inertial range can be also confirmed by the plateau of the local slope. The scaling exponent  is found
 to be $\alpha\simeq 0.37$, which is obtained   by a least square
fitting algorithm.
 It is interesting to note that this value is consistent with  the scaling exponent of the first-order
SFs reported  in other studies,
\cite{Benzi1993b,She1994PRL,Arneodo1996}    indicating almost the same 
intermittent correction on  the probability space and the statistical moments space.
For comparison, the first-order SF $S_1(\tau)=\langle \vert
Y_{\tau} \vert \rangle$ is also shown as $\triangle$. {Note that the absolute value of increments does
not change the result of this paper.} For display convenience, it has been
converted by taking
$1/S_1(\tau)$. It predicts the same inertial range as $p_{\max}(\tau)$. The
corresponding scaling exponent is found to be $\zeta(1)\simeq0.34$, very close to the
Kolmogorov value 1/3. 
 One can find that the $p_{\max}(\tau)$ shows a behavior which seems more linear than $S_1(\tau)$ on the inertial range, see also Fig.\,\ref{fig:maxpdfT} for temperature data. This is because  the first-order SF $S_1(\tau)$ is more sensitive to   large-scale structures, which might pollute the whole inertial range, see  the discussion below and an example of passive scalars with large-scale ramp-cliff structures in Ref. \onlinecite{Huang2010PRE}.

We  note that the inertial range predicted by Fourier power spectrum
$E(f)$ is different from the one predicted by first-order SF $S_1(\tau)$
and $p_{\max}(\tau)$. This phenomenon has been reported by several
authors for the second-order SF $S_2(\tau)$ and the Fourier
power spectrum $E(f)$.
\cite{Nelkin1994,Frisch1995,Hou1998,Huang2009EPL,Huang2010PRE}  If one
considers the Wiener-Khinchin theorem,
\cite{Percival1993,Frisch1995,Huang2010PRE} $S_2(\tau)$ and
 $E(f)$  can be related to each other as 
 \begin{equation}
 S_2(\tau)=\int_0^{+\infty} E(f)\left(1-\cos\left(2\pi \tau f\right)\right) \upd f \label{eq:WKT}
 \end{equation}
 Therefore, for a scaling time series, they are expected to have the
same inertial range,  and the corresponding scaling exponents are related as $\beta=1+\zeta(2)$.
The difference  may come from the following reasons:
(\romannumeral1) the finite  power law range, \cite{Hou1998,Huang2010PRE}
(\romannumeral2) the spectrum of the original velocity is not a pure power law,
\cite{Nelkin1994,Frisch1995} (\romannumeral3) violation of the
statistical
stationary assumption, and  (\romannumeral4) also the
influence of large-scale structures. \cite{Huang2010PRE,Huang2009PHD}  More detail of the discussion can be found in Ref. \onlinecite{Huang2010PRE}. We will
turn to this point again in the next section.

 \subsection{Temperature as an  active scalar from Rayleigh-B\'{e}nard Convection }

 \begin{figure}[!htb]
\begin{center}
\includegraphics[width=.95\linewidth]{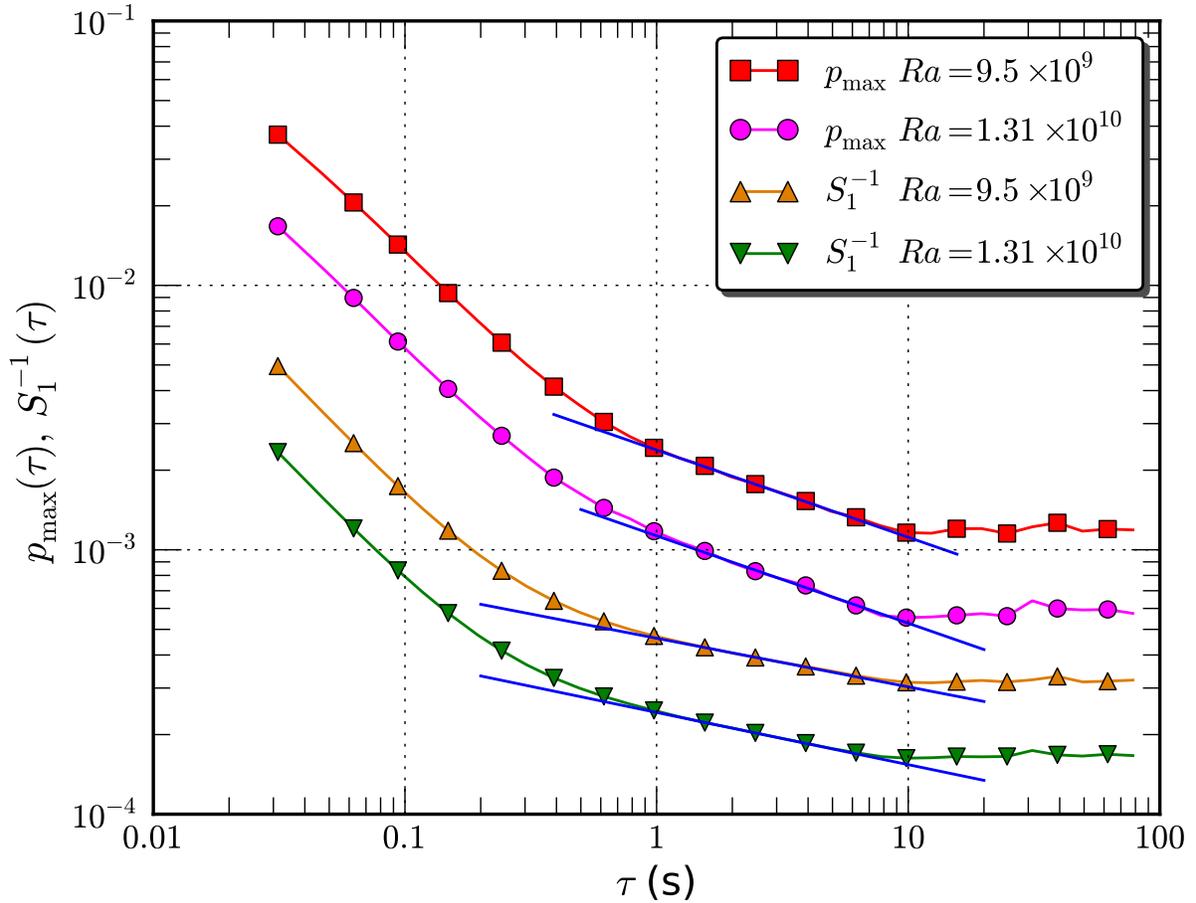}
\caption{  $p_{\max}(\tau)$ and the first-order structure functions
$S_1^{\theta}(\tau)$ of temperature. Power law behavior is found over the range $1<\tau<10\,$s for all
curves. The corresponding scaling exponents are $\alpha_{\theta}\simeq 0.33$
and $\zeta_{\theta}(1) \simeq 0.19$,  respectively. For display convenience,   we have shown the inverse value $1/S_1^{\theta}(\tau)$ of the first-order structure function and the
curves have been
 vertically shifted.  }\label{fig:maxpdfT}
\end{center}
\end{figure}

 \begin{figure}[!htb]
\begin{center}
\includegraphics[width=.95\linewidth]{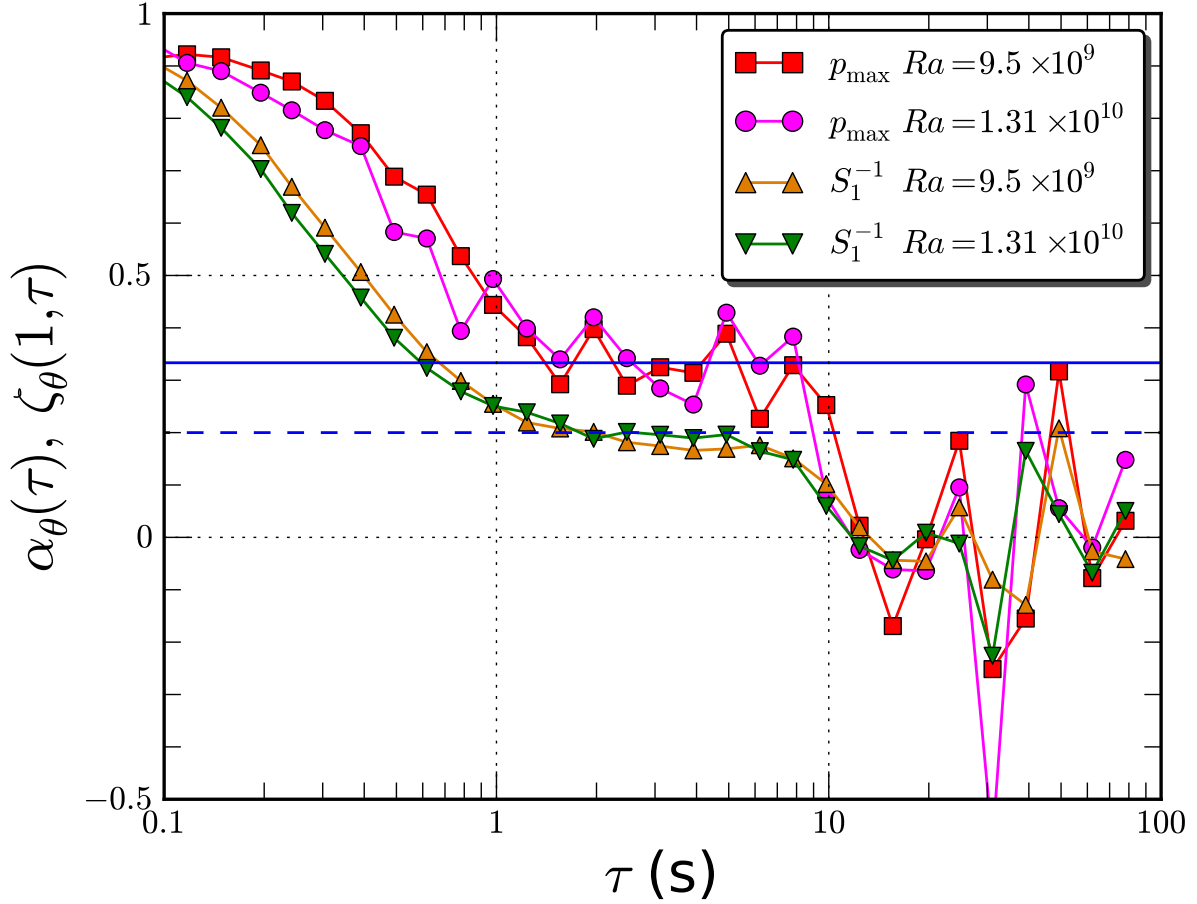}
\caption{  Local slope  for the $p_{\max}(\tau)$ and $S^{-1}_1(\tau)$. The solid horizontal line is the KOC scaling $1/3$, and the dashed line is the BO59 scaling $1/5$.  }\label{fig:TLS}
\end{center}
\end{figure}

We finally consider a temperature data sets obtained near the sidewall of turbulent
RBC system.
 The experiments were performed by Prof. Xia's group in the Chinese University
of Hong Kong. The details of the experiments have been described elsewhere.
\cite{shang2003prl, shang2004pre, Shang2008PRL} Briefly, the temperature
measurements were carried out in a cylindrical cell with upper and lower copper
plates and Plexiglas sidewall. The inner diameter of the cell is $D=19.0$ cm and
the height is $L=19.6$ cm. So its aspect ratio is $\Gamma=D/L\simeq1$. Water was
used as the working fluid and measurement were made at Rayleigh number
$Ra=9.5\times10^{9}$ and $1.31\times10^{10}$. During the experiments, the entire
cell was placed inside a thermostat box whose temperature matches the mean
temperature of the bulk fluid that was kept at $\sim40^{\circ}$C, corresponding
to a Prandtl number $Pr=4.4$. The local temperature was measured at 8 mm
from the sidewall at midheight by using a small thermistor of 0.2 mm diameter
and 15 ms time constant. Typically, each measurement of temperature lasted 20 h
or longer with a sampling frequency 64 Hz, ensuring that the statistical
averaging is adequate.

Near the sidewall of a turbulent convection cell, the turbulent flow is driven
by buoyancy in
the vertical direction. As proposed by Bolgiano and Obukhov, there is a typical
length scale in buoyancy-driven turbulence, now commonly referred to as the
Bolgiano scale $L_B$, above which buoyancy effects are important and the Bolgiano-Obukhov (BO59) 
scaling $E_{\theta}(k)\sim k^{-7/5}$ for temperature power spectrum or
$S_q^{\theta}(r)\sim r^{q/5}$ for SFs are expected.
\cite{Ahlers2009RMP, Lohse2010} Whether the BO59 scaling exists in turbulent
RBC system has been studied extensively in the past two decades, whereas it
remains a major challenge to settle this problem (see, for a recent review, Ref.
\onlinecite{Lohse2010}). Nevertheless, it has been shown recently that above a certain scale buoyancy effects indeed become
predominant, at least in the time domain. \cite{ching2004jot,
zhou2008pre}  The observed Bolgiano time scale
here is of
order 1 second. \cite{zhou2001prl, ching2004jot, zhou2008pre}
 
Figure \ref{fig:maxpdfT} and \ref{fig:TLS} show respectively the estimated $p_{\max}(\tau)$
and the first-order SFs $S_1^{\theta}(\tau)$, and the local slope, in which dashed line indicates the BO59 scaling $1/5$ and the solid line indicates the KOC scaling $1/3$. One can see
the power-law behaviors  or the plateaus  above the Bolgiano time scale, i.e. on the range
$1<\tau<10\,$s. For pdfs, the fitted scaling exponent is $\alpha_{\theta}\simeq
0.33$, which is almost the same as the Kolmogorov-Obukhov-Corrsin (KOC) value of $1/3$  for passive scalar,\cite{Frisch1995,Warhaft2000}
 whereas for SFs, the fitted scaling exponent $\zeta_{\theta}(1)\simeq 0.19$, which is
very close to the BO59 value of $1/5$. At first glance, these results seem to be
contradicting and confusing. To understand this, we note that in turbulent RBC
buoyant forces are exerted on the fluid mainly via thermal plumes. As revealed
by several visualizations,  thermal
plumes consist of a front with sharp temperature gradient and hence these
thermal structures would induce intense temperature increments, which correspond
to the pdf tails. \cite{zhang1997pof, xi2004jfm, zhou2007prl} Therefore, it is not surprising that $p_{\max}(\tau)$
investigated here could not capture efficiently the information of thermal
plumes and thus may preclude buoyancy effects. See next section for more
discussion.

Note that the Taylor's frozen-flow hypothesis,
\begin{equation}
\label{eq:rt}
r_T=-\langle u\rangle\tau,
\end{equation}
is always used to relate the time domain results, such as those shown in Fig.\,\ref{fig:maxpdfT}, to the theoretical predictions made for the space domain. However, the conditions for the Taylor's hypothesis are often not met in turbulent RBC system and hence its applicability to the system is at best doubtful. \cite{sun2006prl, zhou2008jfm, Lohse2010} Recently, based on a second order approximation, He and coworkers \cite{he2006pre, he2009pre} advanced an elliptic model for turbulent shear flows. Later, the model was validated in turbulent RBC system indirectly using the temperature data by Tong and coworkers \cite{tong2010pre, tong2011pre} and directly using the velocity data by \citet{zhou2011jfm}. The most important implication of the elliptic model is that the model can be used to translate time series to space series via
\begin{equation}
\label{eq:re}
r_E=-(U^2+V^2)^{1/2}\tau
\end{equation}
where $U$ is a characteristic convection velocity proportional to the mean velocity and $V$ is a characteristic velocity associated with the r.m.s. velocity and the shear-induced velocity. As pointed by  \citet{zhou2011jfm}, $r$ is proportional to $\tau$ for both the Taylor's relation Eq.\,\eqref{eq:rt} and the elliptic relation Eq.\,\eqref{eq:re}, but   the proportionality constants of the two relations are different. This implies that the Taylor's hypothesis and elliptic model would yield the same scaling exponents. Therefore, if one is only interested in the scaling exponents, one does not really need the validity of Taylor's hypothesis to reconstruct the space series from the measured time series.

\section{Discussion and Conclusion}\label{sec:DCs}

\begin{figure}[!htb]
\begin{center}
\includegraphics[width=1.0\linewidth]{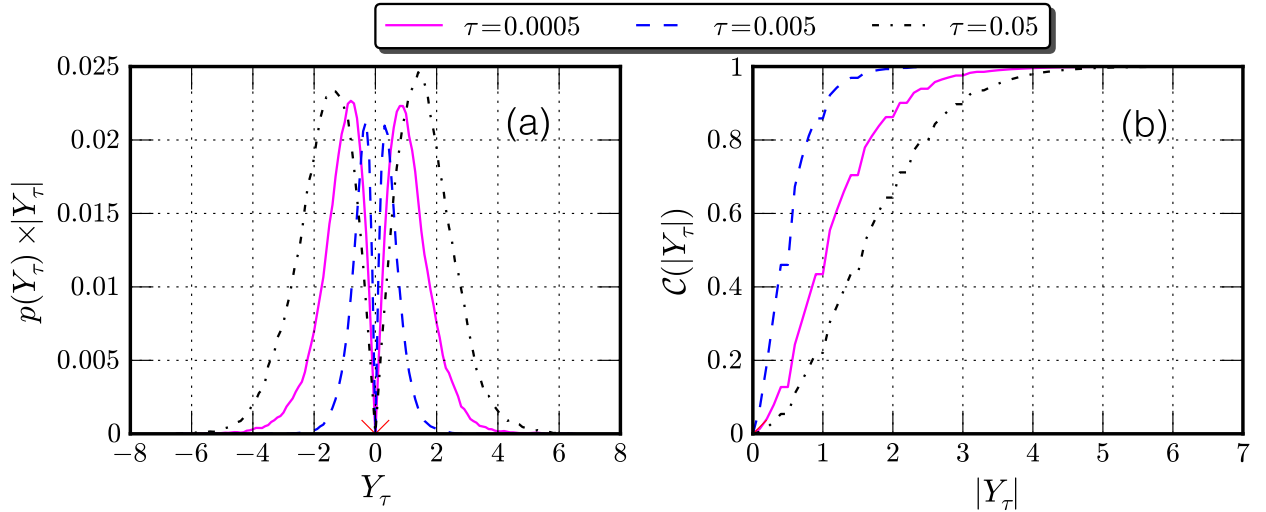}
\caption{(a)  the
integral kernel of first-order structure function, and (b)
the normalized cumulative function $\mathcal{C}(\vert Y_{\tau} \vert)$. The location of $p_{\max}(\tau)$ is
 at $Y_{\tau}\simeq
0$
and marked by
$\times$.}\label{fig:pdfV2}
%\end{minipage}
\end{center}
\end{figure}

\begin{figure}[!htb]
\begin{center}
\includegraphics[width=0.85\linewidth]{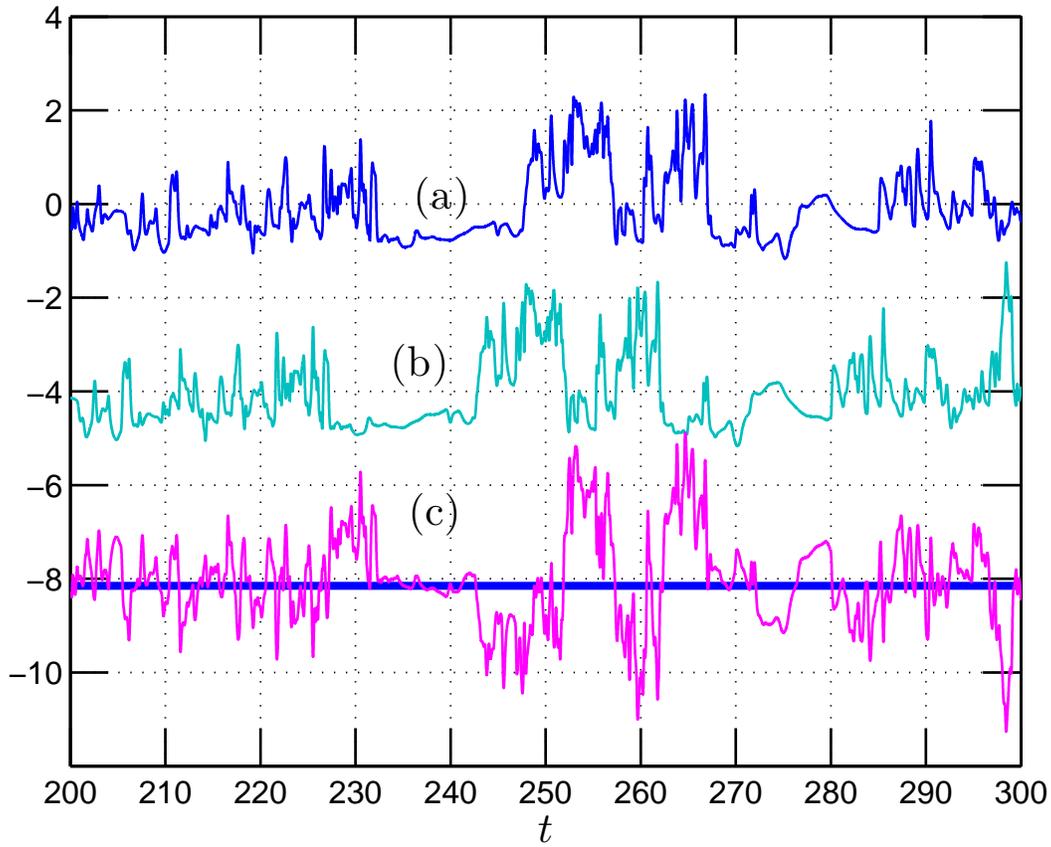}
\caption{Demonstration of the influence of large-scale
structures (plume). They are (a) a portion of temperature data $\theta(t)$,
(b) $\theta (t+\tau)$, and (c) increment
$Y^\theta_{\tau}(t)$ with $\tau=5\,$s, respectively. The $p_{\max}$ is
located
at $Y^ \theta_{\tau}\simeq -0.15$, which is indicated by a small horizontal
patch. For clarity, the curves have been
vertical shifted.}\label{fig:small}
%\end{minipage}
\end{center}
\end{figure}

We have mentioned above that for the turbulent velocity the inertial range predicted by Fourier power
spectrum $E(f)$ is larger than those predicted by the $p_{\max}(\tau)$ and by the first-order
SF.  Indeed,
it has been reported by several authors
that the inertial range predicted by the second-order SF
$S_2(\tau)$ is
shorter than Fourier
power spectrum. \cite{Nelkin1994,Frisch1995,Sreenivasan1997,Hou1998}
By taking
an assumption of statistical stationary and
Wiener-Khinchin theorem, $S_2(\tau)$ and $E(f)$
 can be related with each other,\cite{Percival1993,Frisch1995} see  
Eq.\,\eqref{eq:WKT}.  Thus
 both methods are
expected  to predict an identical inertial range.
\cite{Frisch1995,Sreenivasan1997}
However, the statement of the Wiener-Khichin theorem  only exactly holds for a
stationary process, which may be not satisfied by the turbulent velocity.

As pointed by  \citet{Huang2010PRE}, the second-order SF is also strongly
influenced by the large-scale structures. We show this point experimentally here.
A more rigorous discussion can be found in Ref. \onlinecite{Huang2010PRE}. Figure
\ref{fig:pdfV2} shows (a) the  integral kernel $p(Y_{\tau})\times \vert Y_{\tau} \vert$ 
of the first-order SF for three same separation scales as in Fig.\,\ref{fig:pdfv} (a), and (b) the corresponding normalized cumulative function $\mathcal{C}(\vert Y_{\tau} \vert)$,
respectively.
The location of $p_{\max}(\tau)$ is marked as $\times$. The normalized
cumulative
function $\mathcal{C}(\vert Y_{\tau} \vert)$ is defined as
\begin{equation}
 \mathcal{C}(\vert Y_{\tau} \vert)=\frac{\int_{-Y_{\tau}}^{Y_{\tau} } p(Y'_{\tau}) \vert Y'_{\tau} \vert \upd Y'_{\tau} }
{\int_{-\infty}^{\infty } p(Y'_{\tau}) \vert Y'_{\tau} \vert \upd Y'_{\tau}}
\end{equation}
It characterizes the relative contribution to the first-order SF.
 The location of $p_{\max}$ is found graphically to be $Y_{\tau}\simeq 0$, indicating that at this location there is almost no
contribution to SFs. If we consider the large index value of
$Y_{\tau}$ coming
from large-scale structures,
most of the contribution to SFs comes from them.  The
contribution is found also to be increased with the increase of $\tau$, see
 Fig.\,\ref{fig:pdfV2} (b).
Thus $p_{\max}$ is less influenced by large-scale structures,
	revealing a more accurate  scaling exponent for  $\zeta(1)$.

In the sidewall region of RBC system, the
flow is dominated by plumes. \cite{Ahlers2009RMP,Lohse2010} Figure
\ref{fig:small} shows (a)
a portion of temperature data
$\theta(t)$, (b) $\theta(t+\tau)$, and (c) increment $Y^\theta_{\tau}$
with $\tau=5\,$s, respectively. Due to the presence
of plumes, the shape of pdf $p(Y^\theta_{\tau})$ is asymmetric
(not shown
here). \cite{zhou2008pre} The location of $p_{\max}$ is at $Y^
\theta_{\tau}\simeq-0.15$, which is indicated as a small horizontal patch.  For
SFs, they include contribution from all scale structures. On the
contrary, $p_{\max}$ acts a kind of conditional statistic, in which the
contribution from large-scale structures,  e.g. thermal plumes,   is excluded, see Fig.\,\ref{fig:small} (c).  The large-scale structures here are believed to be thermal plumes.
 Thus the scaling of
$p_{\max}$  may represent the scaling property of the background fluctuation, which is
believed to satisfy KOC scaling.
\cite{Ahlers2009RMP,Lohse2010} Indeed, the KOC scaling for the first-order statistical moment has been found by using a generalized autocorrelation function of  the
 increment, which confirms the idea that in the sidewall region the temperature fluctuation can be considered as a KOC background fluctuation  superposed to BO59 fluctuations (thermal plumes).
 
 This result is compatible with the Grossmann-Lohse (GL) theory,\cite{Grossmann2000JFM,Grossmann2001PRL,Grossmann2002PRE} in which the global thermal dissipation $\epsilon_{\theta}$  is decomposed into the thermal dissipation due to the bulk $\epsilon_{\theta,bulk}$ together with the boundary layer $\epsilon_{\theta,BL}$
\begin{equation}
\epsilon_{\theta}=\epsilon_{\theta,bulk}+\epsilon_{\theta,BL}
\end{equation}
 Later, the GL theory has been modified so that the thermal dissipation $\epsilon_{\theta}$  can be decomposed into 
 the thermal dissipation due to the thermal plumes $\epsilon_{\theta,pl}$ together with the turbulent background   
 $\epsilon_{\theta,bg}$
\cite{Grossmann2004PoF}
\begin{equation}
\epsilon_{\theta}=\epsilon_{\theta,pl}+\epsilon_{\theta,bg}
\end{equation}
in which the contribution from the thermal plumes might be related to the boundary layer. Therefore, in the sidewall region, the thermal dissipation is dominated by the thermal plumes (or boundary layer), see more details in Ref. 
\onlinecite{Grossmann2004PoF}. More recently, this picture has been proofed to be correct at least in the central region of the RBC system  by \citet{Ni2011PRL}. Our result here indicates that in the sidewall region, the turbulent background 
should have contribution to the global thermal dissipation as well as the thermal plumes. Or in other words, the KOC  and BO59 scalings might coexist at least for the temperature fluctuations. We will show this result elsewhere.

 The method we proposed here may be refined by considering some pdf models as 
basis, e.g.  Eq.\,(3.8) in Ref. \onlinecite{Castaing1990}. However, it seems that the Eq.\,(3.8) 
requires the resolution of the spatial dissipation scale $\eta$ to determine a parameter $\sigma_0$, the most 
probable variance of conditional velocity $u$ at a given dissipation rate $\epsilon$, see 
more details in Ref. \onlinecite{Castaing1990}. Unfortunately, the data set we have has no 
resolution on dissipation scale.\cite{Kang2003}  More data sets and pdf models 
will be considered in future studies to refine our method.

  One advantage of the present 
method to consider scaling properties of time series is its ability to exclude the influence of large-scale structure as much as possible.  
Indeed, we have observed a Kolmogorov-like pdf scaling for other data set, in which other 
moment-based methods do not detect the power-law behavior. It is believed that the scaling is destroyed by large-scale structures (result not shown here).

In summary,  we investigated the pdf scaling of velocity increments $Y_{\tau}(t)$. 
 We   postulated a 
scaling relation of the maxima
value of the  pdfs, e.g. $p_{\max}(\tau)\sim \tau^{-\alpha}$.
 We obtained this scaling relation analytically for fBm processes and more generally for $H$-self-similar processes with $\alpha=\zeta(1)$. For the former one, it has been validated by fBm simulations.
The pdf scaling exponent $\alpha$ is comparable with the scaling exponent $\zeta(1)$ of the first-order SFs.
 To our knowledge, at least for $H$-self-similar processes, this is the first method to look at scaling properties on the probability space rather than on the statistical moments space as done classically. We postulated that the pdf scaling holds for   multifractal processes as well. For multifractal processes, due to the failure of Eq.\,\eqref{eq:PR0}, the scaling relation may be only an approximation.
 
When applying this approach to turbulent
velocity, it is found that, statistically speaking, $p_{\max}(\tau)$ satisfies
both normal and lognormal distributions.  A scaling
exponent $\alpha\simeq 0.37$ is found experimentally  which is consistent with the scaling exponent $\zeta(1)$ of the first-order SFs reported in other studies, indicating
 the same intermittent correction on both the probability space and the statistical moments space.  For temperature near the
sidewall of RBC system, the scaling exponent
is found to be $\alpha_{\theta}\simeq0.33$. This value is in favor of KOC scaling  for passive scalar,
not BO59 scaling.  It indicates that  the KOC scaling may be extracted by a proper designed method.
We show experimentally that the  contribution of plumes to
$p_{\max}$ is almost excluded, whereas the SF contains
contributions  from both small-scale  and larger-scale structures. Indeed
the contribution from  the  former ones is much smaller than from the later  ones. Thus
the pdf scaling exponent $\alpha_{\theta}$ represents the scaling property of  the 
background flow.   In other words, the  temperature fluctuation in the side wall region of a RBC can 
be considered as a KOC background fluctuation superposed to BO59 fluctuations (thermal plumes). The first-order KOC scaling exponent has been 
confirmed by using other approach. The potential application of these new findings 
may  serve
as a constrain of some turbulent models,  for example, the pdf model (Eq.\,(3.8)) in Ref. \onlinecite{Castaing1990}.

 \begin{acknowledgments}

This work is sponsored in part by the Key Project of Shanghai Municipal Education Commission (No. 11ZZ87) and the Shanghai Program for Innovative Research Team in Universities and in part by the National Natural Science Foundation of China under
Grant Nos. 10772110 and 11072139. X.S. thanks the financial support from  the National Natural Science Foundation of China under No. 10972229 and U1033002.
 Y. H. thanks Prof. S.Q. Zhou in SCSIO for useful discussion.  We thank Prof. K.-Q. Xia from Chinese University of Hongkong for providing us the temperature data.
We also thank Prof. Meneveau for sharing his experimental
velocity database, which is available for download at
C. Meneveau's web page: http://www.me.jhu.edu/meneveau/datasets.html.
 We also thank the anonymous referees for their useful suggestions.
\end{acknowledgments}

%\bibliography{all}% Produces the bibliography via BibTeX.


\end{document}